\documentclass[sigconf]{acmart}

\usepackage{multirow}
\usepackage{algorithm}
\usepackage{algorithmic}
\usepackage{enumitem}
\usepackage{colortbl}
\usepackage{fontawesome5}

\graphicspath{{Images/}}

\copyrightyear{2026}
\acmYear{2026}
\setcopyright{cc}
\setcctype{by}
\acmConference[CIKM '26]{Proceedings of the 35th ACM International Conference on Information and Knowledge Management}{November 07--11, 2026}{Rome, Italy}
\acmBooktitle{Proceedings of the 35th ACM International Conference on Information and Knowledge Management (CIKM '26), November 07--11, 2026, Rome, Italy}
\acmDOI{10.1145/3799682.3840842}
\acmISBN{979-8-4007-2539-5/2026/11}
\newcommand{\method}{HCGRec}
\newcommand{\sid}{Semantic ID}
\newcommand{\sids}{Semantic IDs}
\newcommand{\grpo}{GRPO}
\makeatletter
\newcommand{\markcorrespondingauthor}{%
  \expandafter\gdef\csname typeset@author\the\num@authors\endcsname##1{%
    ##1\textsuperscript{\faEnvelope}}}
\newcommand{\equalcontributionmark}{\scalebox{1.25}{*}}
\newcommand{\markequalcontributionauthor}{%
  \expandafter\gdef\csname typeset@author\the\num@authors\endcsname##1{%
    ##1\textsuperscript{\equalcontributionmark}}}
\newcommand{\addequalcontributionnote}{%
  \g@addto@macro\@authornotes{%
    \begingroup
    \renewcommand{\thefootnote}{\equalcontributionmark}%
    \footnotetext{These authors contributed equally to this work.}%
    \endgroup}}
\newcommand{\addcorrespondingauthornote}{%
  \g@addto@macro\@authornotes{%
    \begingroup
    \renewcommand{\thefootnote}{\faEnvelope}%
    \footnotetext{Corresponding authors.}%
    \endgroup}}
\makeatother
\definecolor{MethodHighlight}{HTML}{FFF2CC}
\newcommand{\cmark}{\textcolor{green!45!black}{\ensuremath{\surd}}}

\title{Learning from Unreachable Rewards: Hint-Conditioned Reinforcement Learning for Generative Recommendation}

\author{Kangning Zhang}
\markequalcontributionauthor
\email{zhangkangning@sjtu.edu.cn}
\affiliation{%
  \institution{Shanghai Jiao Tong University}
  \city{Shanghai}
  \country{China}
}

\author{Haotian Fang}
\markequalcontributionauthor
\email{fanghaotian@sjtu.edu.cn}
\affiliation{%
  \institution{Shanghai Jiao Tong University}
  \city{Shanghai}
  \country{China}
}

\author{Xukun Luo}
\email{luoxukun@meituan.com}
\affiliation{%
  \institution{Mei Tuan}
  \city{Beijing}
  \country{China}
}

\author{Hao Yin}
\email{yinhao15@meituan.com}
\affiliation{%
  \institution{Mei Tuan}
  \city{Beijing}
  \country{China}
}

\author{Yang Gao}
\email{gaoyang94@meituan.com}
\affiliation{%
  \institution{Mei Tuan}
  \city{Beijing}
  \country{China}
}

\author{Peng Yan}
\email{yanpeng04@meituan.com}
\affiliation{%
  \institution{Mei Tuan}
  \city{Beijing}
  \country{China}
}

\author{Weiwen Liu}
\markcorrespondingauthor
\email{liuweiwen8@huawei.com}
\affiliation{%
  \institution{Shanghai Jiao Tong University}
  \city{Shanghai}
  \country{China}
}

\author{Weinan Zhang}
\markcorrespondingauthor
\email{wnzhang@sjtu.edu.cn}
\affiliation{%
  \institution{Shanghai Jiao Tong University}
  \city{Shanghai}
  \country{China}
}

\author{Yong Yu}
\markcorrespondingauthor
\email{yyu@apex.sjtu.edu.cn}
\affiliation{%
  \institution{Shanghai Jiao Tong University}
  \city{Shanghai}
  \country{China}
}
\addequalcontributionnote
\addcorrespondingauthornote

\renewcommand{\shortauthors}{Zhang et al.}

\begin{document}

\begin{abstract}
Semantic-ID generative recommenders represent each item as a short sequence of discrete semantic tokens and predict the next item by autoregressively generating this token sequence. This paradigm enables a unified generation interface for item IDs, histories, and item text, but it also creates a structured optimization bottleneck during reward-based post-training: when an early semantic token enters the wrong branch of the item-token space, finite rollout groups rarely reach the ground-truth item, so group-relative optimization receives identical zero rewards and produces no useful advantage. We propose \emph{Hint-Conditioned Generative Recommendation} (\method), a semantic-ID generative recommendation framework that recovers learning signal for such hard training instances. \method{} diagnoses each instance with checkpoint rollouts and supplies a minimal target-prefix hint only when the current generator cannot reach the correct item. The model then generates the unhinted suffix under the hinted semantic branch, turning zero-reward groups into informative comparisons over item-token completions. Hinting also changes token identity: hinted prefix tokens are oracle-provided item context, while unhinted suffix tokens are sampled generation actions. We therefore introduce \emph{hint-aware credit decomposition}, using supervised learning to preserve item-semantic and prefix-structure alignment for hinted tokens and \grpo{} to optimize the sampled suffix. Experiments on sequential recommendation benchmarks show that \method{} substantially improves over supervised fine-tuning and vanilla reward-based post-training, while reducing zero-advantage training samples from over 70\% to below 20\%. The code is accessible at \url{https://github.com/WncFht/GRec}.
\end{abstract}



\maketitle


\section{Introduction}
\label{sec:introduction}
Generative recommendation turns item retrieval into language-model-style generation: instead of scoring every candidate item, the model generates the discrete identifier of the next item a user is likely to consume. This formulation is attractive for large-scale sequential recommendation because item identifiers can be organized as short \sid{} token sequences, often learned from item text or multimodal representations through clustering or residual quantization, and a Transformer can then predict the next item by autoregressively decoding these tokens \cite{rajput2023recommendersystemsgenerativeretrieval}. It also creates a natural interface with large language models: interaction histories, item titles, descriptions, and other textual fields can be used to align item identifiers with the model's semantic space \cite{geng2023recommendationlanguageprocessingrlp,zhou2025onerectechnicalreport}.

Recent semantic-ID generative recommenders usually follow a two-stage recipe: supervised semantic alignment followed by reward-based post-training. A model is first supervised on several semantic-alignment tasks, such as predicting the next item's \sid{} from historical \sid{} sequences, from historical title sequences, or from item-side text. It is then optimized with recommendation rewards to align generated items with user preference and business objectives \cite{deng2025onerecunifyingretrieverank,zhou2025onerectechnicalreport,zhou2025onerecv2technicalreport,zou2026genrecpreferenceorientedgenerativeframework}. Group Relative Policy Optimization (\grpo{}) is a common optimizer for this stage because it compares multiple generated item-token sequences from the same prompt and avoids training a separate value model \cite{shao2024deepseekmathpushinglimitsmathematical}. In principle, this converts recommendation feedback into a learning signal over generated \sid{} tokens.

However, the central failure mode is that many semantic-ID training examples become \emph{finite-rollout unreachable} under reward-based generative training. Item-level reward is sparse, but the more important issue is that this sparsity is organized by the \sid{} prefix tree. A generated sequence receives useful signal only when it reaches the ground-truth item or a sufficiently relevant item; if an early token chooses the wrong semantic branch, later tokens are generated under an already incorrect prefix and are unlikely to recover the target item. As a result, under the current generator and a fixed rollout budget, all sampled completions can stay outside the reward-distinguishable region, all rewards are identical, and \grpo{} produces zero relative advantage. In our logged unhinted post-training runs, a large portion of rollout groups remains inactive even late in training, meaning that substantial training data no longer contributes useful group-relative updates.

This paper argues that the obstacle is not merely sparse reward, but \emph{reward reachability within the item-token space}. Reward shaping can provide denser values after a generated sequence is evaluated, and correction-based decoding can repair trajectories after the model has already produced a draft \cite{xing2026learningreflectcorrectbetter}. These strategies are useful, but they do not directly address the training-time question: \textbf{how can a semantic-ID generator learn from an example when its current rollout distribution almost never enters the part of the item-token tree where reward becomes distinguishable?} The structure of semantic item identifiers gives a domain-specific answer: the target \sid{} prefix itself can serve as a minimal scaffold that moves an unreachable training example back into the correct item branch.

We propose \textbf{Hint-Conditioned Generative Recommendation} (\method{}) to recover reachability for hard semantic-ID training examples. Before reward-based post-training, \method{} runs the current checkpoint on each training instance to estimate whether the ground-truth item is reachable under the current generator. If the model can already generate the target item, the instance is trained without modification. If the instance is unreachable, \method{} exposes a short target-prefix hint, such as the first \sid{} token, and asks the model to generate the remaining suffix. The hint is used only as a training-time scaffold for generative recommendation: it does not add side information at inference time, and it is applied only to examples whose unhinted generations fail to produce learning signal.

The key insight is that hinting is \textbf{prefix-tree reachability control}, not answer leakage. A four-token item identifier defines a coarse-to-fine path in the item space. When the generator repeatedly chooses the wrong first token, the suffix-generation problem is ill-posed because all suffixes are conditioned on the wrong branch. Revealing the first token for such hard cases relocates generation to the correct coarse branch, where the remaining tokens still need to be generated by the model and where group-relative rewards can distinguish better and worse item completions. This converts previously inactive examples into active generative training groups while preserving on-policy learning for the unhinted suffix.

\method{} also requires \textbf{hint-aware credit decomposition} inside a hinted semantic-ID generation. Once part of the target \sid{} is inserted into the input, those tokens are no longer sampled actions from the recommender; they are oracle-provided item context. Treating the full \sid{} sequence as a generated action would therefore mix two different sources of information: the hinted prefix did not result from model exploration and should not receive \grpo{} credit, whereas the unhinted suffix is the model's actual generation and should be optimized by \grpo{}. At the same time, the hinted prefix cannot simply be ignored, because it carries item semantics and the coarse-to-fine prefix structure used to locate the correct item branch. We therefore use supervised learning to provide semantic anchoring for the hinted prefix, while \grpo{} optimizes the sampled suffix under the hinted context. This differs from global supervised-plus-reward regularization \cite{zhou2025onerectechnicalreport,zou2026genrecpreferenceorientedgenerativeframework}: the learning signal is assigned according to whether each token is oracle-provided item context or a model-generated item-token action.

Empirically, \method{} improves semantic-ID generation by turning inactive hard examples into active generative training groups. We evaluate \method{} on sequential recommendation benchmarks under the same semantic-index setting as the base generative recommender. The main result is that reachability-aware hinting is competitive with SFT-only training and reward-based generative baselines and improves many key recommendation metrics, especially on deeper ranking cutoffs where stable semantic-branch generation matters. More importantly, it changes the training dynamics by substantially reducing inactive zero-gradient rollout groups during post-training. Ablations further separate the effect of the hinting policy, the training-task scope, and hint-aware credit decomposition, demonstrating that the gains come from targeted reachability recovery rather than simply adding more supervised tokens.

Our contributions are summarized as follows:
\begin{itemize}[leftmargin=*]
    \item We identify \textbf{finite-rollout unreachable reward groups} as a major bottleneck in semantic-ID generative recommendation, where multi-token item identifiers make early-token errors collapse most reward-based training groups to zero advantage.
    \item We introduce \textbf{\method{}}, a hint-conditioned generative recommendation framework that diagnoses unreachable training instances with checkpoint rollouts and supplies minimal target-prefix hints only for hard examples, recovering informative suffix-generation comparisons without changing inference.
    \item We propose \textbf{hint-aware credit decomposition}, which treats hinted prefix tokens as oracle-provided item context and sampled suffix tokens as generated item-token actions, using supervised learning to preserve prefix semantics and \grpo{} to optimize suffix generation.
    \item We show empirically that \method{} improves many key generative recommendation metrics and substantially reduces the inactive-sample ratio on logged post-training runs, providing direct evidence that reachability recovery is critical for semantic-ID generative recommendation.
\end{itemize}


\section{Related Work}
\label{sec:related-work}

\subsection{Sequential and Semantic-ID Generative Recommendation}

Sequential recommendation traditionally formulates next-item prediction as
discriminative ranking over a catalog. GRU4Rec
\cite{hidasi2016sessionbasedrecommendationsrecurrentneural} models
interaction histories with recurrent neural networks, Caser
\cite{tang2018personalizedtopnsequentialrecommendation} uses convolutional
sequence embeddings to capture local sequential patterns, and SASRec
\cite{kang2018selfattentivesequentialrecommendation} and BERT4Rec
\cite{sun2019bert4recsequentialrecommendationbidirectional} establish
self-attention as a strong backbone for user behavior modeling.
Generative recommendation changes the output interface by generating an item
identifier rather than scoring every item. TIGER is a representative
semantic-ID framework \cite{rajput2023recommendersystemsgenerativeretrieval}:
item content representations are quantized into short token sequences, and a
Transformer predicts the next item's \sid{} by autoregressive decoding.
Follow-up studies extend this idea with contrastively learned tree-structured
identifiers \cite{si2024generativeretrievalsemantictreestructured}, unified
generative-dense retrieval
\cite{yang2024unifyinggenerativedenseretrieval}, and practical semantic-ID
pipelines \cite{ju2025generativerecommendationsemanticids}.

Recent work further studies how semantic identifiers should be constructed and
used. CoST \cite{zhu2024costcontrastivequantizationbased} improves semantic
tokenization with contrastive quantization, MTGRec
\cite{zheng2025pretraininggenerativerecommendermultiidentifier} augments
pre-training with multiple identifiers per item, HiD-VAE
\cite{fang2025hidvaeinterpretablegenerativerecommendation} learns
hierarchical and disentangled identifiers, and DIGER
\cite{fu2026differentiablesemanticidgenerative} makes semantic-ID learning
differentiable. Variable-length semantic IDs
\cite{khrylchenko2026variablelengthsemanticidsrecommender} and long semantic
IDs \cite{xia2026unleashpotentiallongsemantic} further revisit the fixed-length
bottleneck of residual quantization.
Other generative-recommendation studies address broader settings, such as
joint generative search and recommendation
\cite{penha2025semanticidsjointgenerative}, cross-task search-recommendation
transfer \cite{penha2024bridgingsearchrecommendationgenerative},
contextualized generative recommendation with SynerGen
\cite{gao2025synergencontextualizedgenerativerecommender}, codebook
rebalancing with CRAB \cite{fan2026crabcodebookrebalancingbias}, and
non-uniform quantization with CArD
\cite{wei2026cardnonuniformquantizationvisual}, and diffusion-based cold-start
item modeling \cite{zhang2026diffcold} and ID-free multimodal token
representations \cite{zhang2024motor}.
These works primarily improve the item-token space itself. \method{} is
orthogonal: given a semantic-ID space, it studies why reward-based
post-training cannot learn from many hard instances and uses target-prefix
hints to restore reachability inside the existing item-token tree.

\subsection{Supervised Semantic Alignment for Recommendation Generation}

Language-model-based recommenders often rely on supervised fine-tuning to
align user histories, item text, and recommendation outputs in a shared
generation space. P5 \cite{geng2023recommendationlanguageprocessingrlp}
formulates recommendation as a family of text-to-text tasks over interactions,
metadata, and reviews. GenRec \cite{ji2023genreclargelanguagemodel} and
TALLRec \cite{bao2023tallreceffectiveefficienttuning} show that large language
models can be adapted to recommendation through instruction-style or
tuning-based supervision. LC-Rec
\cite{zheng2023adaptingcollaborativesemanticsrecommendation} injects
collaborative semantics into large language models and is closely related to
the item-text and sequence fusion tasks used in our supervised stage. In
semantic-ID
systems, this alignment stage usually includes predicting the next \sid{} from
historical \sids{}, predicting identifiers from textual histories, and
mapping item-side text to item identifiers.
Multimodal recommendation likewise studies alignment between content and
ID-based representations \cite{liu2024alignrec}.

Industrial and large-scale generative recommenders adopt a similar SFT-first
recipe. OneRec \cite{deng2025onerecunifyingretrieverank} unifies retrieval and
ranking with a generative recommender and iterative preference alignment. The
OneRec technical report \cite{zhou2025onerectechnicalreport} and GenRec
\cite{zou2026genrecpreferenceorientedgenerativeframework} combine semantic-ID
generation with supervised objectives before reward-based or
preference-oriented post-training.
\method{} assumes this supervised alignment has already produced a reasonable
semantic-ID generator. The problem we target appears after SFT: under
teacher forcing, the model can learn the target token sequence, but under
finite on-policy rollouts it may still fail to enter the correct semantic
branch. Therefore, our contribution is not a new SFT task or a replacement for
semantic alignment. Instead, \method{} uses the SFT checkpoint to diagnose
which instances are rollout-unreachable and then applies minimal target-prefix
hints only to recover reward-based learning signal for those hard cases.

\subsection{Reward-Based Post-Training for Generative Recommendation}

Reward-based post-training aligns generated recommendations with objectives
beyond next-token likelihood. GRPO provides a practical group-relative policy
optimization objective by comparing multiple rollouts from the same prompt and
normalizing their rewards within the group
\cite{shao2024deepseekmathpushinglimitsmathematical}. OneRec
\cite{deng2025onerecunifyingretrieverank} applies iterative preference
alignment, and OneRec-V2 \cite{zhou2025onerecv2technicalreport} further
incorporates real user feedback and duration-aware reward shaping in production
recommendation. GenRec
\cite{zou2026genrecpreferenceorientedgenerativeframework} proposes GRPO-SR,
combining GRPO with NLL regularization and hybrid rewards for
preference-oriented large-scale generative recommendation. MiniOneRec
\cite{kong2025minionerecopensourceframeworkscaling} provides an open-source
framework for scaling generative recommendation experiments, while LoopTool
\cite{zhang2026looptool} and Harness-R1 \cite{shao2026harnessr1} extend
reinforcement learning to model-aware data evolution and executable agent
harness editing.

Several recent methods study how to make reward-optimized generative
recommendation more effective. ReRe
\cite{tan2025reinforcedpreferenceoptimizationrecommendation} introduces
reinforced preference optimization with constrained sampling and auxiliary
ranking rewards, GRC \cite{xing2026learningreflectcorrectbetter} adds a
generation-reflection-correction trajectory, V-STAR
\cite{jiang2026spendsearchpaysvalueguided} uses value-guided structured
sampling and Sibling-GRPO, and Rank-GRPO
\cite{zhu2026rankgrpotrainingllmbasedconversational} changes the credit unit
for conversational recommendation from a whole sequence to rank positions.
These methods recognize that naive sequence-level reward optimization can
produce weak, noisy, or poorly allocated learning signals. \method{} addresses
a different but complementary bottleneck: before correction, ranking, or value
guidance can help, many semantic-ID training instances produce rollout groups
with identical rewards because early tokens enter the wrong item branch. Our
method focuses on making such hard instances reachable under the rollout
budget and then assigning credit only to the actually sampled suffix actions.

\subsection{Hints, Scaffolds, and Credit Assignment under Sparse Rewards}

Our method is also related to hint- or scaffold-based learning under sparse
verifiable rewards. HiLL observes that GRPO can suffer from advantage collapse
when all rollouts in a group receive the same reward, and learns hints to
recover informative groups for hard reasoning tasks
\cite{xia2026learninghintreinforcementlearning}. Scaf-GRPO
\cite{zhang2026scafgrposcaffoldedgrouprelative} similarly studies scaffolded
group-relative optimization for reasoning models. These works show that
external structure can make sparse-reward optimization more effective, but they
are designed for general reasoning rather than semantic-ID item generation.
MMSkills \cite{zhang2026mmskills} provides reusable multimodal procedural
scaffolds for visual agents.

\method{} differs in three ways. First, the hint is not a free-form reasoning
scaffold or an inference-time assistant; it is a target-prefix token sequence
defined by the semantic-ID prefix tree and used only during training. Second,
the hint is selected by an offline reachability diagnosis: if an instance is
already reachable without hints, no target token is exposed. Third, hinting
creates a token-source-specific credit assignment problem
\cite{zhang2026vad}. The hinted prefix is
oracle-provided item context and should be semantically anchored by supervised
loss, whereas the suffix is the model's sampled recommendation action and
should receive group-relative policy credit. This token-source-aware
decomposition is the key distinction between \method{} and the more general
practice of mixing supervised and reward-based losses in generative
recommendation.


\section{Method}
\label{sec:method}

\begin{figure*}[t]
    \centering
    \includegraphics[width=\textwidth]{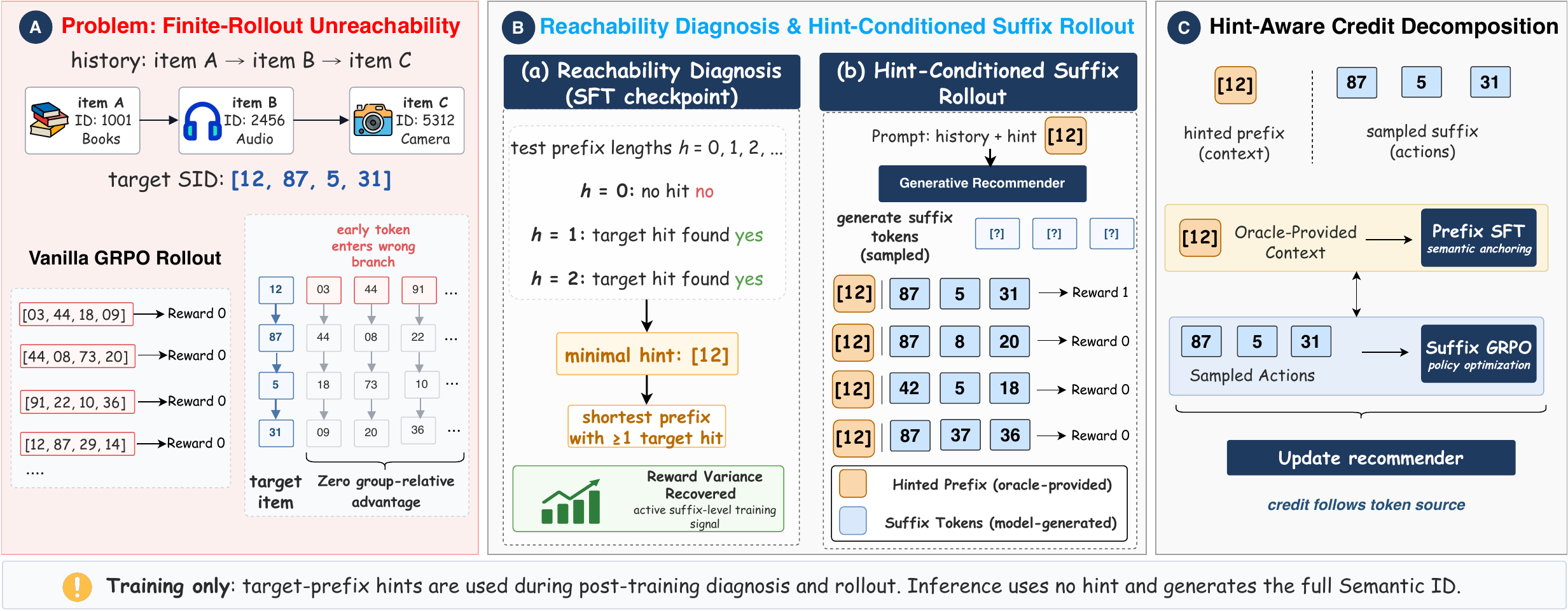}
    \caption{Overview of \method{}. \textbf{(A)} In semantic-ID generative
    recommendation, unhinted rollouts may enter wrong prefix branches and
    produce identical rewards, leaving the instance inactive for group-relative
    optimization. \textbf{(B)} \method{} diagnoses each instance with an SFT
    checkpoint, selects the shortest target-prefix hint that makes the target
    reachable within the rollout budget, and performs suffix rollouts under the
    hinted semantic branch. \textbf{(C)} Hint-aware credit decomposition treats
    the hinted prefix as oracle-provided context anchored by SFT, while the
    sampled suffix remains the model action optimized by \grpo{}.}
    \Description{A pipeline diagram with three parts: finite-rollout
    unreachability under unhinted generation, reachability diagnosis with
    hint-conditioned suffix rollout, and hint-aware credit decomposition for
    semantic-ID generative recommendation.}
    \label{fig:method_overview}
\end{figure*}

This section formalizes \method{} for semantic-ID generative recommendation.
We first introduce the standard semantic-ID generation objective, then show why
finite rollout groups can become reward-unreachable, and finally present
reachability-aware hinting with hint-aware credit decomposition.
Figure~\ref{fig:method_overview} provides an overview of the three central
components of our method.

\subsection{Semantic-ID Generative Recommendation}
\label{sec:method_setup}

Let $\mathcal{I}$ denote the item catalog. Each item $v \in \mathcal{I}$ is
represented by a semantic identifier
$\mathbf{y}(v)=(y_1,\ldots,y_M)$, where $y_t \in \mathcal{V}_{\mathrm{SID}}$ is a
discrete semantic token and $M$ is the identifier length. In the common
residual-quantization setting, $M$ is small; for example, $M=4$ in our
implementation. Given a user context $x$, such as a sequence of historical
items, titles, or item-side textual fields, a generative recommender defines an
autoregressive distribution over the next item's \sid{}:
\begin{equation}
    \pi_\theta(\mathbf{y}\mid x)
    =
    \prod_{t=1}^{M}
    \pi_\theta(y_t \mid x, y_{<t}),
    \label{eq:autoregressive_sid}
\end{equation}
where $\theta$ denotes the model parameters.

The supervised alignment stage trains the model to connect user histories,
item text, and item identifiers in the same generation space. For a training
set $\mathcal{D}=\{(x_i,\mathbf{y}_i^\star)\}$, the basic teacher-forced loss is
\begin{equation}
    \mathcal{L}_{\mathrm{SFT}}(\theta)
    =
    -\mathbb{E}_{(x,\mathbf{y}^\star)\sim\mathcal{D}}
    \sum_{t=1}^{M}
    \log \pi_\theta(y_t^\star \mid x, y_{<t}^\star).
    \label{eq:sft}
\end{equation}
In practice, $x$ can be instantiated by multiple semantic-alignment tasks:
historical \sid{} sequences, title histories, or item-side text. This stage
provides a reasonable generator, but it does not guarantee that on-policy
rollouts can \textbf{reach the target \sid{} path under a finite sampling
budget}.

\subsection{Reward-Based Post-Training and Unreachable Groups}
\label{sec:method_unreachable}

After SFT, reward-based post-training assigns a scalar recommendation reward to
generated item-token sequences. For a prompt $x$, we sample a group of
$G$ candidate identifiers
$\{\hat{\mathbf{y}}_j\}_{j=1}^{G}\sim\pi_{\theta_{\mathrm{old}}}(\cdot\mid x)$
and evaluate each completion with a reward
$r_j=R(x,\hat{\mathbf{y}}_j,\mathbf{y}^\star)$. The reward function can include
exact target match, valid-item constraints, or other recommendation feedback;
\method{} only requires that higher reward means a better generated item.
\grpo{} converts the group rewards into relative advantages:
\begin{equation}
    A_j
    =
    \frac{r_j-\mu_r}{\sigma_r+\epsilon},
    \quad
    \mu_r=\frac{1}{G}\sum_{j=1}^{G}r_j,
    \quad
    \sigma_r^2=\frac{1}{G}\sum_{j=1}^{G}(r_j-\mu_r)^2,
    \label{eq:advantage}
\end{equation}
where $\epsilon$ is a small constant. With the token-level probability ratio
\begin{equation}
    \rho_{j,t}(\theta)
    =
    \frac{
    \pi_\theta(\hat{y}_{j,t}\mid x,\hat{y}_{j,<t})
    }{
    \pi_{\theta_{\mathrm{old}}}(\hat{y}_{j,t}\mid x,\hat{y}_{j,<t})
    },
    \label{eq:ratio}
\end{equation}
and the clipped surrogate
\begin{equation}
    \ell_{\mathrm{clip}}(\rho,A)
    =
    \min\left(
    \rho A,\,
    \mathrm{clip}(\rho,1-\varepsilon,1+\varepsilon)A
    \right).
    \label{eq:clip_surrogate}
\end{equation}
the standard group-relative objective is
\begin{equation}
    \mathcal{L}_{\mathrm{GRPO}}
    =
    -\frac{1}{G}
    \sum_{j=1}^{G}
    \frac{1}{M}
    \sum_{t=1}^{M}
    \ell_{\mathrm{clip}}\left(\rho_{j,t}(\theta),A_j\right).
    \label{eq:vanilla_grpo}
\end{equation}

The difficulty in semantic-ID recommendation is that the reward is structured
by the \textbf{prefix tree of item tokens}. If early tokens choose the wrong coarse
semantic branch, all later tokens are decoded under a wrong prefix and the
generated identifier is unlikely to reach the ground-truth item. We define a
training instance as \textbf{\emph{finite-rollout unreachable}} when its sampled group
does not produce reward-distinguishable completions:
\begin{equation}
    U_\theta(x,\mathbf{y}^\star)
    =
    \mathbb{I}
    \left[
    \mathrm{Var}\left(
    \{R(x,\hat{\mathbf{y}}_j,\mathbf{y}^\star)\}_{j=1}^{G_d}
    \right)
    \leq \tau
    \right],
    \quad
    \hat{\mathbf{y}}_j\sim\pi_\theta(\cdot\mid x),
    \label{eq:unreachable_indicator}
\end{equation}
where $G_d$ is the diagnostic rollout budget and $\tau$ is a small threshold.
For exact-match rewards, $\tau=0$ identifies groups where every completion
receives the same reward, usually all zero. In this case $A_j=0$ for all
rollouts, and Eq.~\eqref{eq:vanilla_grpo} contributes no useful update for that
training instance.

\subsection{Reachability-Aware Hint Conditioning}
\label{sec:method_hinting}

\method{} recovers learning signal by conditioning hard instances on the
\textbf{shortest target-prefix hint} that makes the target item reachable in diagnostic
rollouts. Let $\pi_{\mathrm{sft}}$ be the checkpoint after supervised
semantic alignment. We use this checkpoint once before reward-based
post-training to decide how much prefix information each training instance
needs.

For a target identifier $\mathbf{y}^\star=(y_1^\star,\ldots,y_M^\star)$ and a
candidate hint length $h$, define
\begin{equation}
    \mathbf{p}_h^\star=(y_1^\star,\ldots,y_h^\star),
    \quad
    x_h=[x;\mathbf{p}_h^\star],
    \label{eq:hint_prompt}
\end{equation}
where $\mathbf{p}_0^\star$ is empty and $x_0=x$. Under $x_h$, the model only
needs to generate the remaining suffix. We draw a diagnostic group
\begin{equation}
    \tilde{\mathbf{s}}_{j}^{(h)}
    \sim
    \pi_{\mathrm{sft}}(\cdot \mid x_h),
    \quad
    \tilde{\mathbf{y}}_{j}^{(h)}
    =
    \mathbf{p}_h^\star \oplus \tilde{\mathbf{s}}_{j}^{(h)},
    \quad j=1,\ldots,G_d,
    \label{eq:diagnostic_rollout}
\end{equation}
where $\oplus$ denotes concatenation and $G_d$ is the diagnostic rollout
budget. We say that length $h$ passes the reachability test if at least one
diagnostic completion exactly recovers the target identifier:
\begin{equation}
    B_h(x,\mathbf{y}^\star)
    =
    \mathbb{I}\left[
    \exists j\in\{1,\ldots,G_d\}
    \;\text{s.t.}\;
    \tilde{\mathbf{y}}_{j}^{(h)}=\mathbf{y}^\star
    \right].
    \label{eq:diagnostic_hit}
\end{equation}
Let
\begin{equation}
    \mathcal{C}(x,\mathbf{y}^\star)
    =
    \left\{
    h\in\{0,\ldots,H_{\max}\}: B_h(x,\mathbf{y}^\star)=1
    \right\}
    \label{eq:reachable_hint_set}
\end{equation}
be the set of prefix lengths that make the diagnostic group contain the target
item. The hint used by \method{} is the shortest prefix in this set:
\begin{equation}
    h^\star(x,\mathbf{y}^\star)
    =
    \min \mathcal{C}(x,\mathbf{y}^\star).
    \label{eq:minimal_hint}
\end{equation}
This definition makes the \textbf{hint policy explicit}. If $h^\star=0$, the SFT
checkpoint can already produce the target item without a hint, and the example
is trained in the original form. If $h^\star>0$, the example is hard under
unhinted rollout, and \method{} reveals the \textbf{minimum target prefix} that makes
the diagnostic rollout group contain a correct answer. The guarantee is
therefore finite-budget and checkpoint-specific: the selected prompt has
already produced at least one target hit under $\pi_{\mathrm{sft}}$ and the
diagnostic budget. We choose $H_{\max}<M$ so that at least one semantic token
remains to be generated. If $\mathcal{C}(x,\mathbf{y}^\star)$ is empty under
this budget, the instance is recorded as unresolved in the active-sample
diagnostics; a selected hint is accepted only when it satisfies
$B_{h^\star}(x,\mathbf{y}^\star)=1$.

Given $h=h^\star(x,\mathbf{y}^\star)$, the hinted prompt is $x_h$. During
post-training, the generator samples suffixes
\begin{equation}
    \hat{\mathbf{s}}_{j,h}
    \sim
    \pi_{\theta_{\mathrm{old}}}(\cdot \mid x_h),
    \quad
    \hat{\mathbf{y}}_{j}^{(h)}
    =
    \mathbf{p}_h^\star \oplus \hat{\mathbf{s}}_{j,h},
    \label{eq:hinted_rollout}
\end{equation}
using the same recommendation reward as the unhinted task. When $h=0$,
Eq.~\eqref{eq:hinted_rollout} reduces to the original unhinted generation.
When $h>0$, the hinted prefix relocates the rollout to the correct coarse
branch of the item-token tree, but the remaining suffix is still generated by
the model and evaluated by reward. Thus, the hint is a training-time
\textbf{reachability scaffold}, not inference-time side information.

\subsection{Hint-Aware Credit Decomposition}
\label{sec:method_credit}

Hinting changes which tokens should receive which learning signal. In a hinted
instance, prefix tokens $\mathbf{p}_h^\star$ are oracle-provided item context;
they are not sampled actions from the recommender. The suffix tokens
$\hat{\mathbf{s}}_{j,h}$, in contrast, are the actual generated item-token
actions. \method{} therefore \textbf{decomposes credit by token source}.

For the sampled suffix, we compute group rewards on the completed identifiers
$\hat{\mathbf{y}}_{j}^{(h)}$ and form advantages $A_j^{(h)}$ as in
Eq.~\eqref{eq:advantage}. The suffix probability ratio is
\begin{equation}
    \rho_{j,t}^{(h)}(\theta)
    =
    \frac{
    \pi_\theta(\hat{y}_{j,t}^{(h)}\mid x_h,\hat{y}_{j,h+1:t-1}^{(h)})
    }{
    \pi_{\theta_{\mathrm{old}}}(\hat{y}_{j,t}^{(h)}\mid x_h,\hat{y}_{j,h+1:t-1}^{(h)})
    },
    \quad t>h.
    \label{eq:hinted_ratio}
\end{equation}
The suffix-only group-relative loss is
\begin{equation}
    \mathcal{L}_{\mathrm{suffix}}^{\mathrm{GRPO}}
    =
    -\frac{1}{G}
    \sum_{j=1}^{G}
    \frac{1}{M-h}
    \sum_{t=h+1}^{M}
    \ell_{\mathrm{clip}}\left(
    \rho_{j,t}^{(h)}(\theta),A_j^{(h)}
    \right).
    \label{eq:suffix_grpo}
\end{equation}
No \grpo{} term is applied to the \textbf{hinted prefix}, because those tokens were not
generated by the policy.

The hinted prefix is nevertheless important: it carries item semantics and the
coarse-to-fine structure that identifies the correct item branch. Ignoring it
would weaken the semantic alignment that the SFT stage originally learned. We
therefore add a \textbf{prefix anchoring loss}
\begin{equation}
    \mathcal{L}_{\mathrm{prefix}}^{\mathrm{SFT}}
    =
    -\mathbb{I}[h>0]
    \sum_{t=1}^{h}
    \log \pi_\theta(y_t^\star \mid x,y_{<t}^\star).
    \label{eq:prefix_sft}
\end{equation}
This loss asks the model to keep recognizing the hinted semantic branch from
the original recommendation context, while the suffix loss optimizes the
model's actual generation under that branch.

The final \method{} objective for a training instance is
\begin{equation}
    \mathcal{L}_{\mathrm{HCGRec}}
    =
    \mathcal{L}_{\mathrm{suffix}}^{\mathrm{GRPO}}
    +
    \lambda
    \mathcal{L}_{\mathrm{prefix}}^{\mathrm{SFT}},
    \label{eq:hcgrec_objective}
\end{equation}
where $\lambda$ controls the strength of semantic anchoring. This objective is
different from applying a global supervised regularizer to all tokens. The
\textbf{prefix receives supervised credit} because it is oracle-provided context, and
the \textbf{suffix receives group-relative credit} because it is sampled behavior from
the recommender.

\subsection{Training and Inference}
\label{sec:method_training}

Algorithm~\ref{alg:hcgrec} summarizes the full procedure. \method{} starts from
an SFT checkpoint, performs an offline reachability diagnosis with that
checkpoint, and stores one hint length for each resolved training instance. The
same reward function is used for unhinted and hinted rollouts; the only change
is where generation starts in the semantic-ID prefix tree. At inference time,
no target prefix is available or used. The recommender generates the complete
\sid{} from the user context exactly as in Eq.~\eqref{eq:autoregressive_sid},
so \method{} \textbf{changes training but not the deployment interface}.

\begin{algorithm}[t]
\caption{\method{} Training}
\label{alg:hcgrec}
\begin{algorithmic}[1]
\REQUIRE SFT checkpoint $\pi_{\mathrm{sft}}$, training set $\mathcal{D}$,
reward $R$, diagnostic budget $G_d$, training group size $G$, maximum hint
length $H_{\max}<M$, anchoring weight $\lambda$
\STATE Initialize $\pi_\theta \leftarrow \pi_{\mathrm{sft}}$
\item[] \textbf{Reachability diagnosis}
\FOR{each $(x,\mathbf{y}^\star)\in\mathcal{D}$}
    \FOR{$h=0,\ldots,H_{\max}$}
        \STATE Build $x_h=[x;\mathbf{p}_h^\star]$ and sample $G_d$
        diagnostic suffixes from $\pi_{\mathrm{sft}}(\cdot\mid x_h)$
        \STATE Compute $B_h(x,\mathbf{y}^\star)$ by
        Eq.~\eqref{eq:diagnostic_hit}
    \ENDFOR
    \IF{$\mathcal{C}(x,\mathbf{y}^\star)\neq\emptyset$}
        \STATE Store $h^\star(x,\mathbf{y}^\star)=
        \min\mathcal{C}(x,\mathbf{y}^\star)$ by
        Eq.~\eqref{eq:minimal_hint}
    \ELSE
        \STATE Mark the instance as unresolved for active-sample diagnostics
    \ENDIF
\ENDFOR
\item[] \textbf{Reward-based post-training}
\FOR{each minibatch $\mathcal{B}\subset\mathcal{D}$}
    \FOR{each resolved $(x,\mathbf{y}^\star)\in\mathcal{B}$}
        \STATE Let $h=h^\star(x,\mathbf{y}^\star)$ and build
        $x_h=[x;\mathbf{p}_h^\star]$
        \STATE Sample $G$ suffixes from
        $\pi_{\theta_{\mathrm{old}}}(\cdot\mid x_h)$
        \STATE Complete each identifier, compute rewards and group advantages
        \STATE Compute $\mathcal{L}_{\mathrm{suffix}}^{\mathrm{GRPO}}$
        by Eq.~\eqref{eq:suffix_grpo} and
        $\mathcal{L}_{\mathrm{prefix}}^{\mathrm{SFT}}$ by
        Eq.~\eqref{eq:prefix_sft}
    \ENDFOR
    \STATE Update $\theta$ with the minibatch average of
    $\mathcal{L}_{\mathrm{HCGRec}}$ in Eq.~\eqref{eq:hcgrec_objective}
\ENDFOR
\end{algorithmic}
\end{algorithm}


\section{Experiments}
\label{sec:experiments}

We organize the experiments around four research questions. These questions examine the overall effectiveness of \method{}, clarify how prefix hints recover useful optimization signal, and identify the contributions of training-task design and hint-aware credit decomposition.

\begin{itemize}[leftmargin=*]
    \item \textbf{RQ1: Overall performance.} Does \method{} improve semantic-ID generative recommendation over supervised training and vanilla reward-based post-training?
    \item \textbf{RQ2: Deep analysis for prefix hint.} Does prefix hinting recover effective training signal, and how does the proposed offline minimal hint differ from dynamic hinting?
    \item \textbf{RQ3: Ablation study.} How do training tasks and loss decomposition affect \method{}?
    \item \textbf{RQ4: Hint-aware credit decomposition weight.} How does the prefix anchoring weight $\lambda$ in Eq.~\eqref{eq:hcgrec_objective} affect recommendation performance across datasets?
\end{itemize}

\subsection{Experimental Setup}
\label{sec:exp_setup}

\subsubsection{Datasets}
We evaluate the method on three public real-world benchmarks from the Amazon Product Reviews dataset \citep{ni2019justifying}: \textit{Musical Instruments}, \textit{Arts, Crafts and Sewing}, and \textit{Video Games}. For brevity, we refer to these three datasets as \textit{Instruments}, \textit{Arts}, and \textit{Games}, respectively. Following the GenRec pipeline \citep{ji2023genreclargelanguagemodel,zou2026genrecpreferenceorientedgenerativeframework}, each item is represented by a four-token \sid{}, and all variants share the same semantic index, backbone, and evaluation pipeline. User behavior sequences are constructed in chronological order, and the maximum history length is set to 50 for all three datasets.

\begin{table*}[t]
\centering
\caption{Statistics of the three evaluation domains and the resulting
sample counts for the training tasks. \textit{SeqRec} reports
train/valid/test counts; \textit{item2index+index2item}, \textit{fusion-seqrec},
\textit{title-seqrec}, and \textit{title/desc2index}
report train-only counts. The last two columns summarize the total
numbers of SFT and RL training samples, respectively.
Sparsity is computed as $1 - \#Interactions / (\#Users \times \#Items)$.}
\label{tab:dataset_statistics}
\scriptsize
\setlength{\tabcolsep}{3pt}
\renewcommand{\arraystretch}{1.05}
\resizebox{\textwidth}{!}{%
\begin{tabular}{lrrrrrccccccc}
\toprule
\multicolumn{6}{c}{Raw Data} & \multicolumn{7}{c}{Training Samples} \\
\cmidrule(lr){1-6}\cmidrule(lr){7-13}
Dataset & \#Users & \#Items & \#Interactions & Sparsity & Avg.~len. & \shortstack{seqrec\\tr/va/te} & \shortstack{item2index\\+index2item} & fusion-seqrec & title-seqrec & \shortstack{title/desc\\2index} & SFT total & RL total \\
\midrule
Instruments & 17,112 & 6,250 & 136,226 & 99.87\% & 7.96 & 84,890 / 17,112 / 17,112 & 12,465 & 84,890 & 10,000 & 11,551 & 182,245 & 106,441 \\
Arts & 22,171 & 9,416 & 174,079 & 99.92\% & 7.85 & 107,566 / 22,171 / 22,171 & 18,743 & 107,566 & 10,000 & 17,061 & 233,875 & 134,627 \\
Games & 42,259 & 13,839 & 373,514 & 99.94\% & 8.84 & 246,737 / 42,259 / 42,259 & 27,287 & 246,737 & 10,000 & 23,373 & 520,761 & 280,110 \\
\bottomrule
\end{tabular}%
}
\end{table*}

Table~\ref{tab:dataset_statistics} summarizes both the raw corpus scale and the resulting task-wise sample counts for the three domains.

\subsubsection{Data Split and Evaluation Metrics}
We adopt the leave-one-out strategy for evaluation. The most recent item in each user sequence is used for testing, the second most recent item is used for validation, and the remaining interactions are used for training. We report top-$K$ Hit Ratio $\left(\textrm{HR@}K\right)$ and Normalized Discounted Cumulative Gain $\left(\textrm{NDCG@}K\right)$ with $K \in \left\{5, 10, 50\right\}$. Evaluation is conducted with full ranking over the \textbf{entire item set} rather than sampled negatives. For the generative methods, the beam size is fixed at 50. For each variant, one checkpoint is selected by $\textrm{NDCG@}10$ from all synchronized checkpoints of that variant, and all reported metrics are read from that same checkpoint.

\subsubsection{Training Tasks}
The data pipeline instantiates five task families. \textit{SeqRec} is the basic sequential recommendation task: given the historical interaction sequence, the model predicts the next item \sid{}. \textit{item2index+index2item} is an item-side semantic alignment task that maps between item titles and semantic identifiers in both directions. \textit{fusion-seqrec} conditions on the historical \sid{} sequence but predicts the title of the next item instead of its semantic identifier. \textit{title-seqrec} uses the title sequence of historical interacted items to predict the next item \sid{}. \textit{title/desc2index} is an item identification task that predicts the target \sid{} from either an item title or an item description.

The supervised stage uses \textit{SeqRec}, \textit{item2index+index2item}, and \textit{fusion-seqrec}. Reward-based post-training starts from the same SFT checkpoint in all comparisons, and its training pool is built from \textit{SeqRec}, \textit{title-seqrec}, and \textit{title/desc2index}. In other words, RL keeps the sequential next-\sid{} objective from \textit{SeqRec}, adds title-sequence-based next-\sid{} prediction through \textit{title-seqrec}, and includes title/description-to-\sid{} grounding through \textit{title/desc2index}, while \textit{item2index+index2item} and \textit{fusion-seqrec} remain SFT-only objectives.

\subsubsection{Reward Function}
For reward-based post-training, we use an \textbf{Exact Target Match Reward}. Given a target semantic identifier $\mathbf{y}^{\star} = (y^{\star}_1,\ldots,y^{\star}_L)$ and a generated semantic identifier $\hat{\mathbf{y}} = (\hat{y}_1,\ldots,\hat{y}_L)$, the rollout reward is
\begin{equation}
    r(\hat{\mathbf{y}}, \mathbf{y}^{\star}) =
    \mathbb{I}\left[\hat{\mathbf{y}} = \mathbf{y}^{\star}\right],
\end{equation}
where the reward is 1 only when the generated \sid{} exactly matches the ground-truth target \sid{}, and 0 otherwise. For hinted rollouts, the hinted prefix is treated as oracle-provided context, and the exact-match check is applied to the complete target \sid{} after concatenating the hint and the generated suffix.

\subsubsection{Implementation Details}
We initialize the generative recommender from Qwen2.5-3B-Instruct and perform full-parameter supervised fine-tuning. Unless otherwise noted, supervised training uses a maximum sequence length of 512, bf16 precision, a cosine learning-rate schedule, learning rate $3\times10^{-4}$, 8 GPUs, per-device batch size 32, and gradient accumulation over 8 steps. Reward-based post-training uses 8 processes with the exact target match reward, per-device train/eval batch size 64, gradient accumulation over 2 steps, learning rate $10^{-5}$, 2 training epochs, temperature 1.0, maximum completion length 128, and 16 sampled completions per prompt. The offline diagnostic pass uses beam size 16, maximum hint depth 3, and default unsolved depth 3. The prefix-anchoring coefficient is set to 0.005 unless otherwise stated.

\subsection{RQ1: Overall Recommendation Performance}
\label{sec:exp_main}

We compare \method{} with representative sequential recommendation baselines, including Caser~\citep{tang2018personalizedtopnsequentialrecommendation}, GRU4Rec~\citep{hidasi2016sessionbasedrecommendationsrecurrentneural}, BERT4Rec~\citep{sun2019bert4recsequentialrecommendationbidirectional}, SASRec~\citep{kang2018selfattentivesequentialrecommendation}, TIGER~\citep{rajput2023recommendersystemsgenerativeretrieval}, and LC-Rec~\citep{zheng2023adaptingcollaborativesemanticsrecommendation}. We also include three reward-based generative baselines, namely \textit{GRPO Rule-only}, \textit{MiniOneRec}~\citep{kong2025minionerecopensourceframeworkscaling}, and \textit{HCGRec (offline hint)}. For a controlled post-training comparison, \textit{GRPO Rule-only}, \textit{MiniOneRec}, \textit{HCGRec (offline hint)}, and \method{} all start from the same SFT checkpoint and use the same semantic index, rollout group size, reward function, decoding constraints, training budget, and checkpoint selection rule. Results are reported in Table~\ref{tab:overall_results}. The compared methods are summarized below.

\begin{itemize}[leftmargin=*]
    \item \textbf{Caser}~\citep{tang2018personalizedtopnsequentialrecommendation} is a convolutional sequential recommender that models local patterns in recent user interactions with horizontal and vertical convolutional filters.
    \item \textbf{GRU4Rec}~\citep{hidasi2016sessionbasedrecommendationsrecurrentneural} is a recurrent sequential recommendation baseline that encodes the interaction history with gated recurrent units and scores candidate next items from the hidden state.
    \item \textbf{BERT4Rec}~\citep{sun2019bert4recsequentialrecommendationbidirectional} uses bidirectional Transformer encoding and masked item prediction, providing a strong non-generative sequential modeling baseline.
    \item \textbf{SASRec}~\citep{kang2018selfattentivesequentialrecommendation} is a self-attentive sequential recommender that models long-range dependencies with causal attention and predicts the next item by item scoring.
    \item \textbf{TIGER}~\citep{rajput2023recommendersystemsgenerativeretrieval} represents the semantic-ID generative retrieval paradigm, in which items are represented as discrete semantic tokens and recommendation is formulated as identifier generation.
    \item \textbf{LC-Rec}~\citep{zheng2023adaptingcollaborativesemanticsrecommendation} adapts language-model-style supervision to collaborative semantics and serves as the supervised semantic-alignment checkpoint in the present pipeline.
    \item \textbf{GRPO Rule-only} is a rule-reward variant of generative recommendation that directly optimizes full generated semantic identifiers with the exact target match reward, without prefix hinting or hint-aware credit decomposition.
    \item \textbf{MiniOneRec}~\citep{kong2025minionerecopensourceframeworkscaling} is a reward-based generative recommendation baseline that follows the OneRec-style post-training recipe with semantic identifiers.
    \item \textbf{HCGRec (offline hint)} is the offline minimal-hint post-training variant that reuses the offline diagnosed minimal reachable prefix during reward-based optimization, but does not apply the hint-aware credit decomposition used by the full method.
    \item \textbf{\method{}} is our method, which combines reachability-aware prefix hinting with hint-aware credit decomposition during post-training.
\end{itemize}

\begin{table*}[t]
\centering
\caption{Overall results on Instruments, Arts, and Games. Bold and underline
mark the best and second-best results; highlight marks the full \method{}.}
\label{tab:overall_results}
\scriptsize
\setlength{\tabcolsep}{3pt}
\renewcommand{\arraystretch}{0.95}
\resizebox{\textwidth}{!}{%
\begin{tabular}{llccccccccc>{\columncolor{MethodHighlight}}c}
\toprule
\multicolumn{2}{c}{}
& \multicolumn{4}{c}{Sequential Rec.}
& \multicolumn{2}{c}{Generative SFT}
& \multicolumn{4}{c}{Reward-based Generative Rec.} \\
\cmidrule(lr){3-6}\cmidrule(lr){7-8}\cmidrule(lr){9-12}
Dataset & Metric & Caser & GRU4Rec & BERT4Rec & SASRec & TIGER & LC-Rec &
\shortstack{GRPO\\Rule-only} & MiniOneRec & \shortstack{HCGRec\\(offline hint)} &
\cellcolor{MethodHighlight}\method{} \\
\midrule
\textit{Gen.} & & -- & -- & -- & -- & \cmark & \cmark & \cmark & \cmark & \cmark & \cmark \\
\textit{SID} & & -- & -- & -- & -- & \cmark & \cmark & \cmark & \cmark & \cmark & \cmark \\
\textit{RL} & & -- & -- & -- & -- & -- & -- & \cmark & \cmark & \cmark & \cmark \\
\midrule
\multirow{6}{*}{Instruments} & HR@5 & 0.0645 & 0.0793 &
0.0790 & 0.0684 & 0.0872 & 0.0932 & \textbf{0.1027} & 0.1002 &
\underline{0.1014} & 0.1009 \\
 & HR@10 & 0.0825 & 0.0880 & 0.0937 & 0.0869 &
0.1044 & 0.1094 & 0.1179 & 0.1145 & \textbf{0.1189} & \underline{0.1180} \\
 & HR@50 & 0.1593 & 0.1291 & 0.1541 & 0.1657 &
0.1763 & 0.1844 & 0.1681 & 0.1696 & \underline{0.1941} & \textbf{0.1985} \\
 & NDCG@5 & 0.0533 & 0.0735 & 0.0689 & 0.0568 &
0.0771 & 0.0771 & \textbf{0.0911} & \underline{0.0906} & 0.0875 & 0.0890 \\
 & NDCG@10 & 0.0591 & 0.0763 & 0.0735 & 0.0628 &
0.0825 & 0.0823 & \textbf{0.0960} & \underline{0.0952} & 0.0931 & 0.0945 \\
 & NDCG@50 & 0.0757 & 0.0850 & 0.0865 & 0.0799 &
0.0980 & 0.0985 & 0.1070 & 0.1071 & \underline{0.1094} & \textbf{0.1118} \\
\midrule
\multirow{6}{*}{Arts} & HR@5 & 0.0421 & 0.0682 & 0.0563 &
0.0593 & 0.0905 & 0.0949 & 0.1031 & 0.1018 &
\underline{0.1041} & \textbf{0.1048} \\
 & HR@10 & 0.0547 & 0.0812 & 0.0734 & 0.0814 &
0.1154 & 0.1164 & 0.1215 & 0.1189 & \underline{0.1248} & \textbf{0.1257} \\
 & HR@50 & 0.1103 & 0.1428 & 0.1431 & 0.1627 &
\textbf{0.2076} & 0.1941 & 0.1906 & 0.1839 & \underline{0.2007} & 0.2006 \\
 & NDCG@5 & 0.0347 & 0.0598 & 0.0461 & 0.0426 &
0.0725 & 0.0801 & \textbf{0.0893} & 0.0876 & \underline{0.0890} & 0.0889 \\
 & NDCG@10 & 0.0387 & 0.0640 & 0.0516 & 0.0498 &
0.0805 & 0.0870 & \underline{0.0952} & 0.0931 & \textbf{0.0956} & \textbf{0.0956} \\
 & NDCG@50 & 0.0508 & 0.0772 & 0.0666 & 0.0673 &
0.1006 & 0.1038 & 0.1102 & 0.1071 & \textbf{0.1120} & \underline{0.1118} \\
\midrule
\multirow{6}{*}{Games} & HR@5 & 0.0334 & 0.0358 & 0.0465 &
0.0340 & 0.0489 & 0.0503 & 0.0544 & 0.0541 &
\underline{0.0555} & \textbf{0.0558} \\
 & HR@10 & 0.0546 & 0.0574 & 0.0723 & 0.0556 &
0.0806 & 0.0804 & \underline{0.0825} & 0.0802 & \textbf{0.0857} & \textbf{0.0857} \\
 & HR@50 & 0.1517 & 0.1621 & 0.1875 & 0.1571 &
\textbf{0.2118} & 0.1998 & 0.1815 & 0.1804 & 0.1972 & \underline{0.2012} \\
 & NDCG@5 & 0.0218 & 0.0232 & 0.0304 & 0.0218 &
0.0318 & 0.0336 & 0.0377 & 0.0374 & \underline{0.0383} & \textbf{0.0385} \\
 & NDCG@10 & 0.0286 & 0.0301 & 0.0387 & 0.0287 &
0.0419 & 0.0433 & \underline{0.0467} & 0.0458 & \textbf{0.0480} & \textbf{0.0480} \\
 & NDCG@50 & 0.0496 & 0.0526 & 0.0636 & 0.0505 &
0.0704 & 0.0691 & 0.0683 & 0.0676 & \underline{0.0723} & \textbf{0.0732} \\
\bottomrule
\end{tabular}%
}
\end{table*}

Based on these comprehensive results, we make the following observations.

\textbf{\method{} delivers its clearest gains where semantic reachability matters most.} Table~\ref{tab:overall_results} shows that \method{} is not uniformly best on every dataset and cutoff. On Instruments, it gives the best $\textrm{HR@}50$ (0.1985) and $\textrm{NDCG@}50$ (0.1118), while the rule-reward and offline-hint variants remain stronger on several top-rank metrics. On Arts and Games, \method{} is strongest or tied for strongest on the headline $\textrm{HR@}5$/$\textrm{HR@}10$ and $\textrm{NDCG@}10$ metrics, but TIGER still leads on $\textrm{HR@}50$. The consistent pattern is therefore more specific than a blanket win: \method{} improves many key metrics while leaving a few extreme operating points to specialized baselines.

\textbf{The comparison with HCGRec (offline hint) separates the effect of reachability correction from the effect of credit decomposition.} Once an offline reachable prefix is introduced, performance already becomes competitive, which confirms that reachability correction is the primary intervention. The full method further improves over the offline-hint-only variant on several metrics, including $\textrm{HR@}50$ and $\textrm{NDCG@}50$ on Instruments, $\textrm{HR@}5$/$\textrm{HR@}10$ on Arts, and $\textrm{HR@}5$/$\textrm{NDCG@}50$ on Games. At the same time, offline hinting remains equal or better on some other cutoffs. This mixed but structured pattern supports the credit-decomposition argument without overstating it: token-source-aware optimization is helpful, but it is not a universal improvement on every metric.

\textbf{The remaining weaknesses are structurally consistent with the design of the method.} The cases in which \method{} does not dominate are concentrated in two settings: very early ranks on Instruments, where the rule-reward variant remains slightly more aggressive, and the deepest cutoff on Arts and Games, where TIGER still performs best. These residual gaps do not contradict the method. Instead, they indicate that reachability-aware hinting mainly improves branch stability and suffix optimization under the correct semantic prefix, which naturally favors overall ranking quality and broad coverage more than every extreme operating point. This is why the empirical profile of \method{} is most convincing as a robust generative recommender rather than as a specialized optimizer for a single cutoff.

\subsection{RQ2: Deep Analysis for Prefix Hint}
\label{sec:exp_prefix_hint}

RQ2 addresses two questions about prefix hints. The first question is whether hinting genuinely recovers useful training signal in group-relative post-training. The second question is whether the proposed offline minimal-hint policy is preferable to dynamic hinting, which computes hint depth during training.

\begin{figure}[t]
    \centering
    \includegraphics[width=\columnwidth]{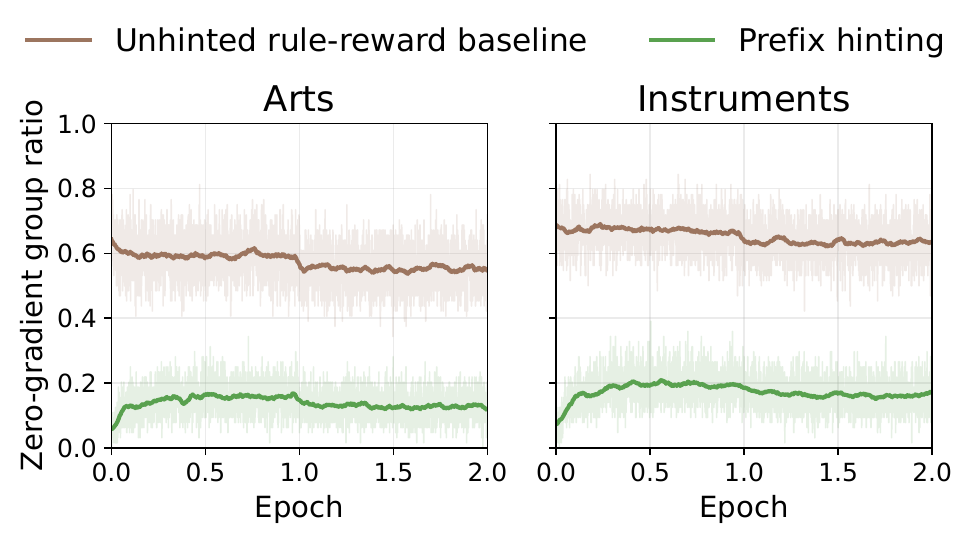}
    \caption{Fraction of rollout groups without gradient signal during training. The y-axis reports the proportion of sampled groups with zero reward variance, where lower values indicate fewer inactive samples.}
    \Description{Line charts for Arts and Instruments comparing unhinted and
    prefix-hinted training. Prefix hinting lowers the zero-gradient group ratio
    throughout training and finishes far below the unhinted baseline on both
    datasets.}
    \label{fig:zero_variance_ratio}
\end{figure}

We quantify the first question with the \textbf{zero-gradient group ratio}, that is, the fraction of training instances whose sampled rollout group satisfies $\operatorname{Var}\left(\left\{r_j\right\}_{j=1}^{G}\right) = 0$, where $G$ denotes the rollout-group size. Such instances yield identical group rewards and therefore do not provide useful group-relative learning signal. As shown in Figure~\ref{fig:zero_variance_ratio}, the unhinted post-training baseline leaves a large fraction of rollout groups in this inactive regime: the smoothed ratio ends around 0.55 on Arts and 0.63 on Instruments. Prefix hinting reduces the corresponding end-of-training ratios to about 0.12 and 0.17, respectively, by moving hard instances onto a reachable semantic branch before suffix rollout. This observation directly supports the central claim that prefix hints are not a superficial input modification. They convert many otherwise inactive samples into active reward-bearing training cases.

\begin{figure}[t]
    \centering
    \includegraphics[width=\columnwidth]{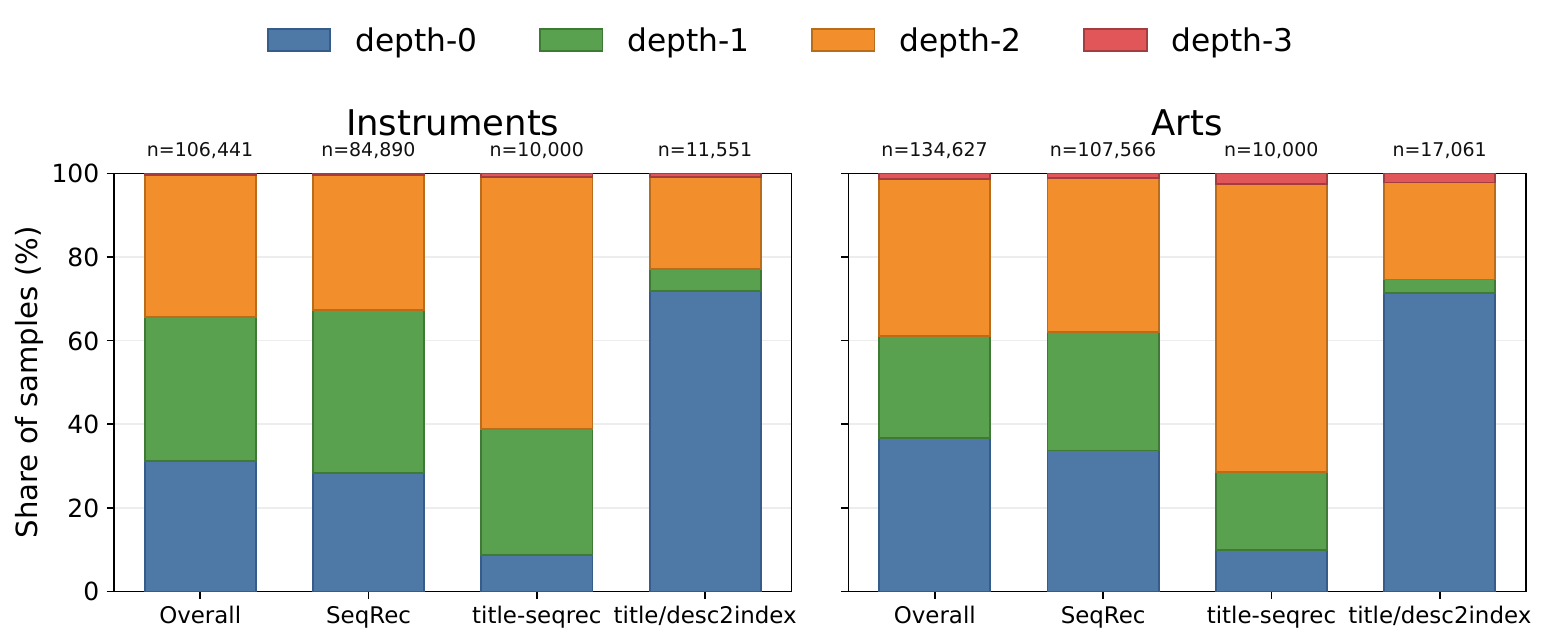}
    \caption{Task-wise fixed-hint depth distributions on Instruments and Arts. Each bar reports the relative depth composition of one RL task and of the overall RL mixture.}
    \Description{Stacked bar charts showing the proportions of fixed hint
    depths for each reinforcement-learning task and for the overall task
    mixture on Instruments and Arts. Title-sequence recommendation contains
    more deep hints, while title- and description-to-index tasks contain mostly
    shallow hints.}
    \label{fig:hint_depth_map_instruments_arts}
\end{figure}

\begin{figure}[t]
    \centering
    \includegraphics[width=\columnwidth]{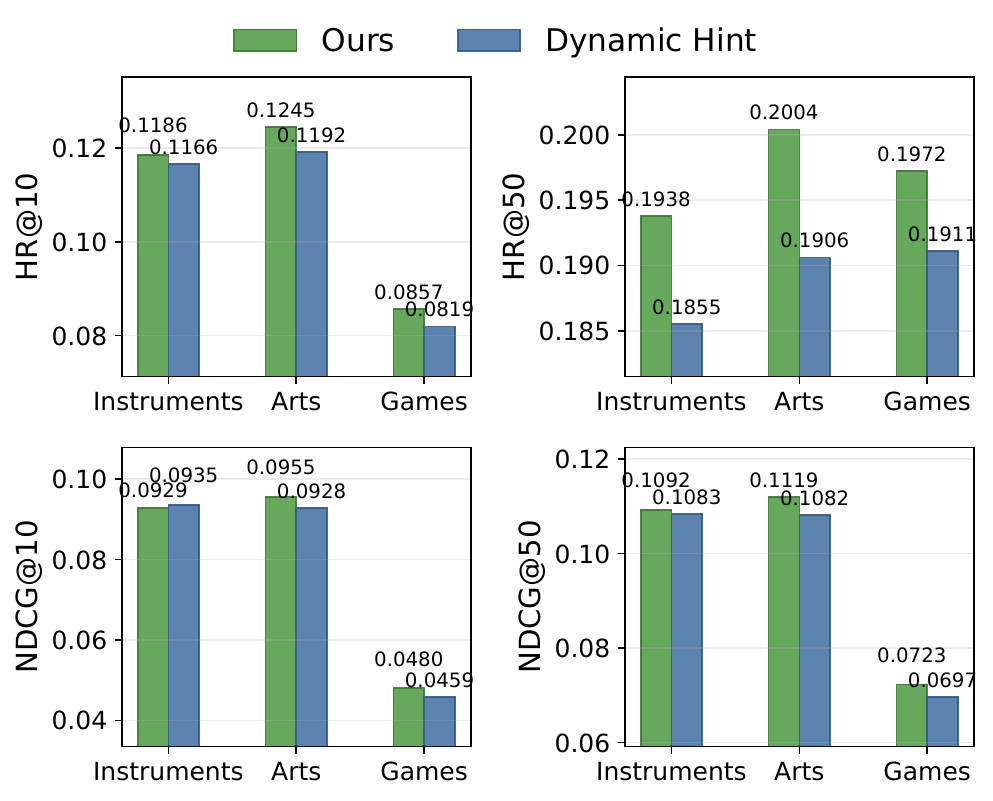}
    \caption{Hint-strategy comparison on Instruments, Arts, and Games. Each panel corresponds to one metric, and each panel reports grouped bars for the three datasets under the same readout rule.}
    \Description{Grouped bar charts comparing the offline minimal-hint policy
    with dynamic hinting across ranking metrics on Instruments, Arts, and
    Games. The offline policy is stronger on most displayed dataset-metric
    combinations.}
    \label{fig:hint_strategy_bars}
\end{figure}

For the second question, we compare the proposed strategy with dynamic hinting using the cross-dataset visualization in Figure~\ref{fig:hint_strategy_bars}. \textbf{Ours} denotes the reachability-aware minimal prefix hint from Section~\ref{sec:method_hinting}: the hint length is selected once by SFT-checkpoint diagnosis and then fixed during reward-based post-training. Dynamic Hint instead recomputes the hint depth online during training.

\textbf{The task-level depth distributions explain why prefix hinting does not affect every RL task equally.} Figure~\ref{fig:hint_depth_map_instruments_arts} shows that the offline hint policy is distributed very differently across tasks. In both datasets, \textit{title-seqrec} places much more mass on deeper hints than \textit{SeqRec}, whereas \textit{title/desc2index} is dominated by shallow cases. This means that the gain from prefix hinting is not only dataset-dependent but also task-dependent: some RL tasks already become reachable after shallow correction, while others rely much more heavily on deeper semantic anchoring before suffix optimization becomes informative.

\textbf{The main advantage of the offline minimal-hint policy is that it keeps optimization focused on a stable target.} Figure~\ref{fig:hint_strategy_bars} shows that dynamic hinting can remain competitive and even slightly better on a few top-heavy metrics, but the offline hint strategy is stronger on most displayed metrics once the comparison is extended to broader cutoffs and multiple datasets. A natural interpretation is that dynamic hinting keeps redirecting training toward whichever hard cases appear most difficult at the current stage. This can preserve isolated short-range gains, but it also makes the optimization objective drift with the model and increases the risk of over-adapting to unstable hard examples. By contrast, the offline policy fixes the semantic decision once and then optimizes the suffix under that fixed branch, which produces a more consistent training signal at lower rollout cost.

\subsection{RQ3: Ablation Study}
\label{sec:exp_ablation}

We further examine task scope and loss decomposition on Instruments.

\textbf{Under the offline minimal-hint setting with prefix anchoring, the full RL task setting becomes the strongest option.} The full task setting, which combines \textit{SeqRec}, \textit{title-seqrec}, and \textit{title/desc2index}, achieves the best $\textrm{HR@}10$ (0.1180), $\textrm{HR@}50$ (0.1985), $\textrm{NDCG@}10$ (0.0945), and $\textrm{NDCG@}50$ (0.1118), while remaining competitive on the other reported metrics. The two-task variant, \textit{SeqRec + title/desc2index}, is generally weaker than the other two settings, with only a negligible advantage over SeqRec-only on $\textrm{HR@}10$.

\begin{figure}[t]
    \centering
    \includegraphics[width=\columnwidth]{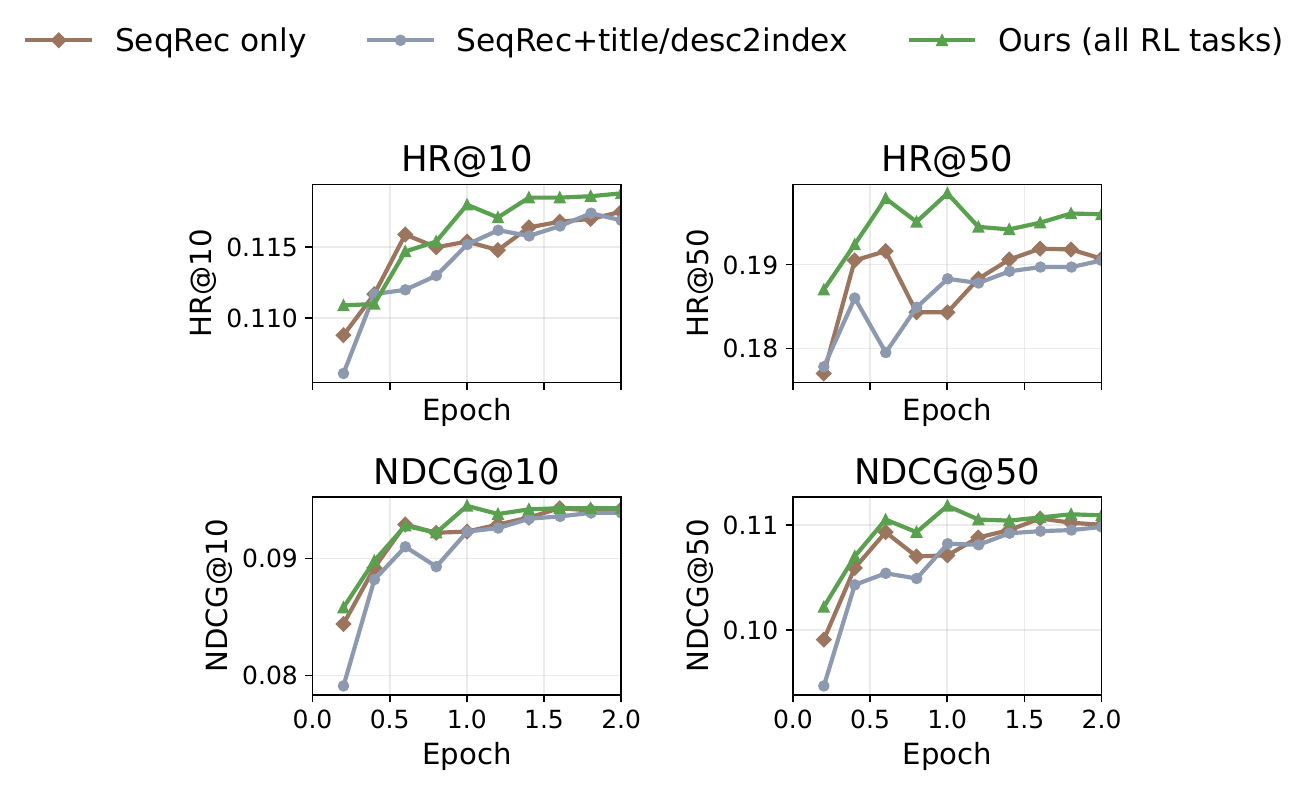}
    \caption{Task-scope trajectories on Instruments under the offline minimal-hint setting with prefix anchoring.}
    \Description{Four training-trajectory plots for hit rate and normalized
    discounted cumulative gain on Instruments. The full reinforcement-learning
    task setting reaches the strongest final values, while the two-task variant
    is generally weakest.}
    \label{fig:task_scope_curves}
\end{figure}

\textbf{Figure~\ref{fig:task_scope_curves} confirms the same training pattern.} Across $\textrm{HR@}10$, $\textrm{HR@}50$, $\textrm{NDCG@}10$, and $\textrm{NDCG@}50$, the full-task line reaches the strongest final checkpoint values, whereas the two-task variant remains the weakest trajectory for most of the run.

\textbf{The task-scope ablation suggests that the full RL task setting provides better generalization than narrower task subsets.} The key point is not only that the full-task setting gives the strongest endpoint results, but also that the additional task coverage improves how consistently the model uses the hinted semantic branch. Once hinting restores reachability, broader supervision helps the model preserve that benefit across different recommendation conditions instead of specializing too strongly to a single task view. In this sense, the main contribution of the full-task setting is better generalization rather than a narrow gain on one isolated metric.

\textbf{The task-scope curves reinforce the same interpretation.} The advantage of the full-task setting is not a single late checkpoint effect; it remains visible across the training trajectory on the headline metrics. This behavior indicates that the extra tasks do not merely add optimization noise or task complexity. Instead, they act as a useful regularizer that improves the generalization of the hinted model and makes the resulting training dynamics more stable.

For loss decomposition, we compare suffix-only \grpo{}, full-sequence SFT + \grpo{}, and our prefix-SFT + suffix-only \grpo{} objective under the same checkpoint-selection rule.

\textbf{\textit{Ours (prefix SFT + suffix-only \grpo{})} remains the strongest loss design.} It achieves the best $\textrm{HR@}5$ (0.1009), $\textrm{HR@}50$ (0.1985), $\textrm{NDCG@}5$ (0.0890), and $\textrm{NDCG@}50$ (0.1118), which indicates that a limited amount of supervision on the hinted prefix is sufficient to improve semantic alignment before reward-based optimization acts on the suffix.

\textbf{The comparison suggests that too much SFT is not beneficial.} Full-sequence SFT + \grpo{} does not improve over the prefix-only design, which means that extending supervised pressure to the entire sequence does not help once the key alignment problem has already been resolved at the prefix level. The more effective strategy is to apply SFT only to the hinted prefix, where semantic alignment is needed, and then let suffix-level policy optimization focus on the sampled continuation.

\textbf{The main gain therefore comes from aligning only the part of the sequence that actually needs alignment.} The hinted prefix serves as semantic context and should be stabilized by supervised learning, whereas the suffix should remain the target of policy optimization. The strongest result comes precisely from preserving this division, rather than from increasing the amount of supervision everywhere in the sequence.

\subsection{RQ4: Effect of Hint-Aware Credit Decomposition Weight}
\label{sec:exp_credit_weight}

We finally study the weight $\lambda$ of hint-aware credit decomposition in Eq.~\eqref{eq:hcgrec_objective} on Instruments and Arts. This coefficient controls how strongly the hinted prefix is anchored with supervised semantic credit relative to suffix-only \grpo{} optimization.

\begin{figure}[t]
    \centering
    \includegraphics[width=\columnwidth]{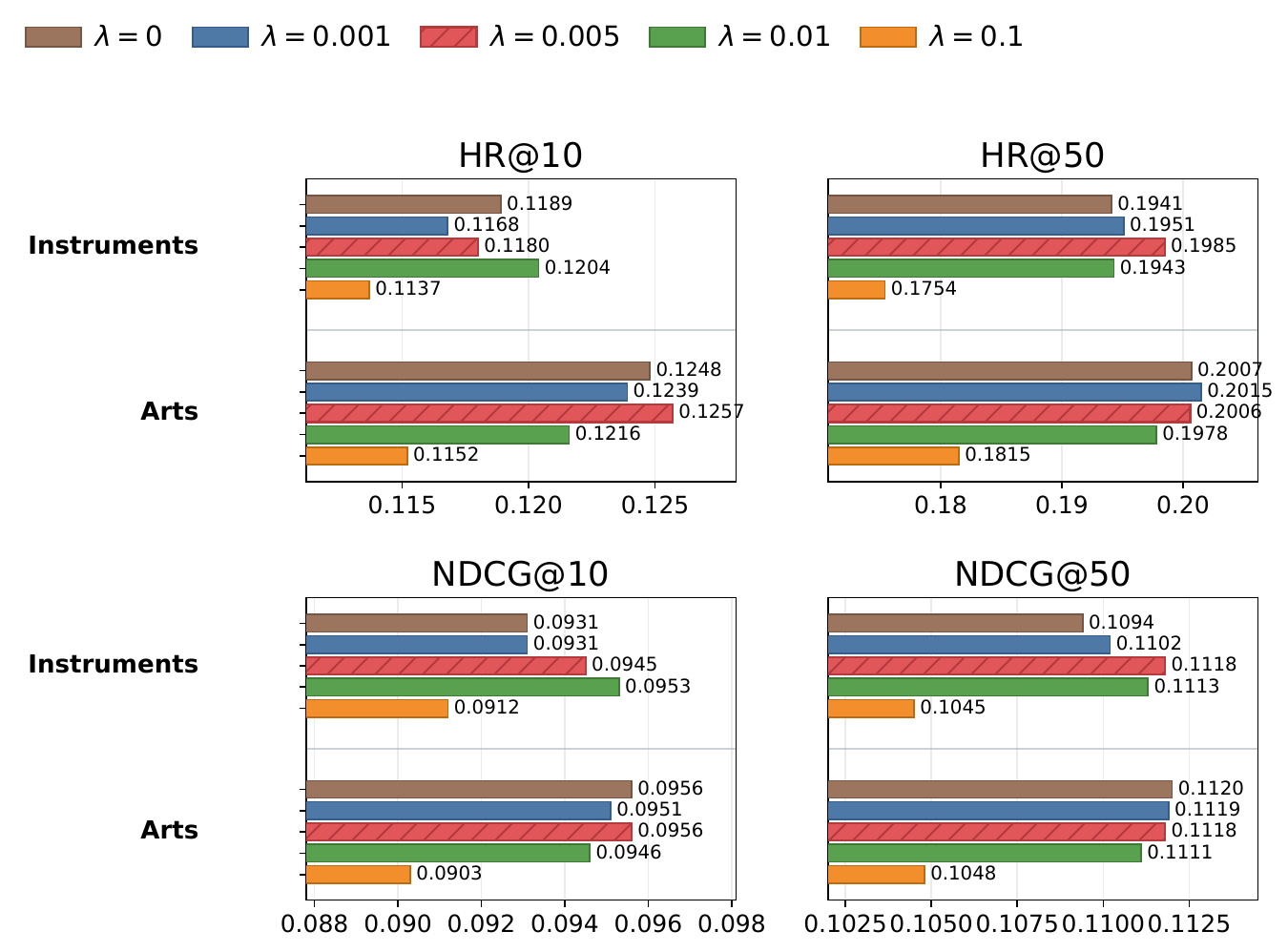}
    \caption{RQ4 credit-weight comparison on Instruments and Arts. Each panel reports one ranking metric, and each dataset group compares $\lambda=0$, $0.001$, $0.005$, $0.01$, and $0.1$. The hatched bar indicates the default \method{} setting used in the main comparison.}
    \Description{Bar charts for four ranking metrics on Instruments and Arts,
    comparing five hint-aware credit-decomposition weights. Moderate weights
    from 0.001 to 0.01 achieve the strongest or near-strongest results, whereas
    the largest weight of 0.1 is consistently weaker.}
    \label{fig:credit_weight_bars}
\end{figure}

\textbf{RQ4 shows that hint-aware credit decomposition is useful, but only as a calibrated bias.} Figure~\ref{fig:credit_weight_bars} shows that increasing $\lambda$ does not monotonically improve performance. Moderate values around $0.001$--$0.01$ produce the strongest or near-strongest results on most metrics, while the overly large setting $\lambda=0.1$ consistently weakens recommendation quality. This behavior is exactly the tradeoff implied by Eq.~\eqref{eq:hcgrec_objective}: the hinted prefix should receive enough supervised anchoring pressure to preserve semantic stability, but not so much that suffix-level reward optimization is overwhelmed.

\textbf{This makes $\lambda$ a calibration knob rather than a scaling knob.} The goal is not to maximize supervised prefix anchoring in isolation, but to calibrate the balance between semantic stability and suffix discrimination. The pattern in Figure~\ref{fig:credit_weight_bars} therefore supports the broader claim of the paper: what matters is not simply to add more supervision after hinting, but to add the right amount in a way that preserves the division of labor between prefix and suffix optimization.


\section{Conclusion}
\label{sec:conclusion}

\method{} addresses finite-rollout reward unreachability in semantic-ID
generative recommendation through minimal target-prefix hints and hint-aware
credit decomposition. Experiments on three Amazon domains show consistent
recommendation gains and substantially fewer zero-variance rollout groups.
Future work will explore online reachability estimation.

\begin{acks}
The work is supported by National Natural Science Foundation of China (62502310).
\end{acks}

\section*{GenAI Usage Disclosure}
Generative AI tools were used to assist with manuscript drafting and language editing. All technical claims, experimental results, citations, and final text are reviewed and verified by the authors.

\bibliographystyle{ACM-Reference-Format}
\bibliography{references}

\end{document}